\documentclass[conference]{IEEEtran}
\IEEEoverridecommandlockouts
\def\BibTeX{{\rm B\kern-.05em{\sc i\kern-.025em b}\kern-.08em
    T\kern-.1667em\lower.7ex\hbox{E}\kern-.125emX}}

\newif\ifarxiv
\arxivtrue %

\newif\ifisanonymized
\isanonymizedfalse  %
\ifarxiv
    \pdfoutput=1
    \usepackage{hologo}
    \usepackage{listings}
\fi

\usepackage[acronyms,toc,nonumberlist,nogroupskip]{glossaries}
\glsdisablehyper
\glsnoexpandfields

\usepackage{verbatim}
\usepackage{tabularx,calc}
\usepackage{color}
\usepackage{adjustbox}
\usepackage{hyperref}
\usepackage{ifthen}
\usepackage{lipsum}
\usepackage{xstring}
\usepackage{amssymb}
\usepackage{amsmath}
\usepackage{siunitx}
\usepackage{graphicx}
\usepackage{bm}
\usepackage{booktabs}
\usepackage{multirow}
\usepackage{booktabs}
\usepackage{colortbl}
\usepackage{microtype}
\usepackage{here}
\usepackage{tabularx}
\usepackage{arydshln}
\usepackage{titlesec}
\usepackage{array}
\usepackage{siunitx}
\usepackage{listings}
\usepackage{adjustbox}
\usepackage{titlesec}
\usepackage{caption}
\usepackage{xspace}
\usepackage{subcaption}
\usepackage{listings}
\usepackage{multicol}
\usepackage[bibstyle=ieee,
			citestyle=numeric-comp,
			backend=bibtex,
			minnames=1,
            maxnames=3,
			maxcitenames=1,
			maxbibnames=4,
			doi=false,
			isbn=false,
			url=false,
			natbib=true]
			{biblatex}

\DefineBibliographyStrings{english}{andothers={et\addabbrvspace al\adddot}}
\DefineBibliographyStrings{english}{phdthesis = {PhD Thesis,}}
\AtEveryBibitem{\clearlist{language}}
\newcommand{\inq}[1]{``#1''}

\newcommand{\verticalCustomWhitespaceHalf}{\vspace{0.075em}}
\renewenvironment{quote}
{\vspace{0.15em}\list{}{\rightmargin=10pt
    \leftmargin=10pt}%
  \item\relax}
{\endlist\vspace{0.15em}}

\newcounter{definitionCounter}

\newcounter{rquestionCounter}

\newcounter{problemCounter}

\newcommand{\requirement}[3]{%
    \par %
    \verticalCustomWhitespaceHalf %
    \begin{quote} %
        \noindent\textit{#2:} #3
        \label{#1}
    \end{quote}
    \verticalCustomWhitespaceHalf %
}

\newcommand{\mobile}{\emph{MOBILE}}

\renewcommand{\vec}[1]{\mathbf{#1}} %

\newcommand{\tm}{\,\mathrm{m}}

\newcommand{\toom}{\mathrm{m}^{-1}} %

\newcommand{\mss}{\,\frac{\mathrm{m}}{\mathrm{s}^2}}
\newcommand{\tmss}{\,\mathrm{m} / \mathrm{s}^2}
\newcommand{\kmh}{\frac{\mathrm{km}}{\mathrm{h}}}
\newcommand{\tkmh}{\,\mathrm{km} / \mathrm{h}}

\newcommand{\odd}{q}

\newcommand{\wheel}{w}%
\newcommand{\degradation}{\mathrm{D}}

\newcommand{\zdegwheel}{\tilde{z}_{\degradation, \wheel}}
\newcommand{\deltadegwheel}{\tilde{\delta}_{\degradation, \wheel}}
\newcommand{\deltadotdegwheel}{\dot{\tilde{\delta}}_{\degradation, \wheel}}
\newcommand{\taudegwheel}{\tilde{\tau}_{\degradation, \wheel}}

\newcommand{\kmin}{k_{\odd}^{\mathrm{min}}}
\newcommand{\kmax}{k_{\odd}^{\mathrm{max}}}
\newcommand{\wmin}{w_{\odd}^{\mathrm{min}}}
\newcommand{\wmax}{w_{\odd}^{\mathrm{max}}}
\newcommand{\amax}{a_{\odd, \mathrm{max}}}

\newcommand{\inputnn}{\xi} %
\newcommand{\inputnntest}{\inputnn_{\mathrm{Test}}}
\newcommand{\inputnnset}{\mathcal{X}}
\newcommand{\inputnntestset}{\inputnnset_{\mathrm{Test}}}
\newcommand{\inputnntrainset}{\inputnnset_{\mathrm{Train}}}
\newcommand{\inputnncalset}{\inputnnset_{\mathrm{Cal}}}

\newcommand{\outputnn}{y}
\newcommand{\outputnntest}{\outputnn_{\mathrm{Test}}}

\newcommand{\interval}{\mathcal{I}}
\newcommand{\suup}{\mathrm{Hi}}
\newcommand{\sulow}{\mathrm{Lo}}
\newcommand{\epsNmax}{\varepsilon_{\mathrm{lat}, \mathrm{max}}}
\newcommand{\epsCutoff}{\varepsilon_{\mathrm{lat}, \mathrm{max}}^{\mathrm{cut}}}

\newcommand{\prob}{P}
\newcommand{\coverage}{C_{\alpha}}

\newcommand{\smallerfootnote}{\fontsize{7}{8.5}\selectfont}
\newcolumntype{C}[1]{>{\centering\arraybackslash}p{#1cm}}

\newcommand{\ads}{automated driving system\xspace}

\newcommand{\aib}{AI-based\xspace}

\newcommand{\aibss}{AI-based systems\xspace}

\newacronym{asoa}{ASOA}{Automotive Service-Oriented Architecture}

\newcommand{\autotech}{autotech.agil\xspace}

\newcommand{\eendarchs}{end-to-end AI-based architectures\xspace}

\newacronym{fmea}{FMEA}{Failure Mode and Effects Analysis}

\newacronym{gsn}{GSN}{Goal Structuring Notation}

\newacronym{hazop}{HAZOP}{Hazard and Operability Analysis}

\newacronym{icd}{ICD}{Interface Control Document}

\newacronym{icds}{ICDs}{Interface Control Documents}

\newacronym{json}{JSON}{JavaScript Object Notation}

\newcommand{\meters}{\,\text{m}\xspace}

\newcommand{\modarch}{modular architecture\xspace}

\newcommand{\modarchs}{modular architectures\xspace}

\newcommand{\mrm}{minimal risk maneuver\xspace}

\newacronym{pilog}{PI Log}{performance indicator log}

\newacronym{pi}{PI}{performance indicator}

\newcommand{\qemph}[1]{\emph{#1}}

\newacronym{ros}{ROS~2}{Robot Operating System 2}

\newcommand{\sawa}{self-awareness\xspace}

\newacronym{servsa}{ServSA}{Service Self-Assessment}

\newacronym{soa}{SOA}{service-oriented architecture}

\newacronym{spi}{SPI}{Safety Performance Indicator}

\newacronym{spis}{SPIs}{Safety Performance Indicators}

\newacronym{stpa}{STPA}{System-Theoretic Process Analysis}

\newacronym{syssa}{SysSA}{System Self-Assessment}

\newcommand{\CQR}{conformalized quantile regression}
\usepackage[none]{hyphenat}
\titleformat{\paragraph}[runin]{\normalfont\normalsize\itshape}{\theparagraph}{0.5em}{\hspace{0.5em}}[:]
\titlespacing*{\paragraph}{0pt}{1ex plus 0.5ex minus 0.2ex}{0.5em}
\renewcommand{\theparagraph}{\alph{paragraph})}

\newcommand{\OffsetTitleToAuthors}{-0.0cm}
\newcommand{\OffsetAuthorsToAbstract}{-1.4cm}

\newcommand{\AnonymizedAuthorsHeight}{0.5cm}

\newcommand{\AnonymizedGrantInfoHeight}{2.1cm}
\newcommand{\AnonymizedGrantInfoExtraSpace}{0.4cm}
\newcommand{\AnonymizedAcknowledgementHeight}{0.9cm}
\ifisanonymized
\IEEEaftertitletext{\vspace{\OffsetAuthorsToAbstract}\vspace{\AnonymizedGrantInfoExtraSpace}}
\else
\IEEEaftertitletext{\vspace{\OffsetAuthorsToAbstract}}
\fi

\newcommand{\titlebaselineskip}{0.25\baselineskip}
\titlespacing*{\section}{-0.75em}{\dimexpr\titlebaselineskip}{\titlebaselineskip}

\newcommand{\eqskipInPt}{5pt}
\AtBeginDocument{%
  \abovedisplayskip=\eqskipInPt
  \abovedisplayshortskip=\eqskipInPt
  \belowdisplayskip=\eqskipInPt
  \belowdisplayshortskip=\eqskipInPt
}

\titleformat{\paragraph}[runin]{\normalfont\normalsize\itshape}{\theparagraph}{0.5em}{\hspace{0.5em}}[:]
\titlespacing*{\paragraph}{0pt}{1ex plus 0.5ex minus 0.2ex}{0.5em}
\renewcommand{\theparagraph}{\alph{paragraph})}
\newcommand{\AuthorOneName}{Ole Reuter$^{1, \dagger}$}

\newcommand{\AuthorTwoName}{Richard Schubert$^{1, \dagger}$}

\newcommand{\AuthorThreeName}{Marvin Loba$^1$}

\newcommand{\AuthorFourName}{Markus Maurer$^1$}

\newcommand{\insertFourAuthors}{%
\author{%
\ifisanonymized
\rule{\textwidth}{\AnonymizedAuthorsHeight}
\else
\IEEEauthorblockN{\AuthorOneName, \AuthorTwoName, \AuthorThreeName, and \AuthorFourName}\\
\vspace{1em}
\fi
}%
}

\newcommand{\titlestring}{Equalized Coverage in Motion Control Performance Prediction for Self-Adaptive Road Vehicles*}

\title{\titlestring%
    \vspace{\OffsetTitleToAuthors}
    \ifisanonymized
    \thanks{\hspace{-1em}\rule{\dimexpr\linewidth}{\AnonymizedGrantInfoHeight}}
    \else
    \thanks{$^{*}$\,This research is partly accomplished within the project \autotech (FKZ~01IS22088R)~\parencite{van_kempen_autotechagil_2023}. We acknowledge the financial support by the Federal Ministry of Research, Technology and Space (BMFTR; former Federal Ministry of Education and Research of Germany, BMBF).}%
	\thanks{$^{1}$\,All authors are with the Institute of Control Engineering, Technische Universität Braunschweig, 38106 Braunschweig, Germany\\ \{o.reuter, richard.schubert, m.loba, markus.maurer\}@tu-braunschweig.de}%
    \thanks{$\dagger$\,These authors contributed equally to this work.}%
    \fi
}

\insertFourAuthors

\begin{document}

\ifarxiv
	\twocolumn[
\begin{@twocolumnfalse}
	\Huge {IEEE copyright notice}
	
	\vspace{0.5cm}
	
	\large {\copyright\ 2026 IEEE. Personal use of this material is permitted. Permission from IEEE must be obtained for all other uses, in any current or future media, including reprinting/republishing this material for advertising or promotional purposes, creating new collective works, for resale or redistribution to servers or lists, or reuse of any copyrighted component of this work in other works.}
	
	\vspace{0.5cm}
	
	{\Large Accepted to be published in \emph{2026 IEEE 29th International Conference on Intelligent Transportation Systems (ITSC)}, Naples, Italy, September 15-18, 2026} 
	
	\vspace{0.75cm}
	
	Cite as:
	\vspace{0.2cm}
	
	\noindent\fbox{%
		\parbox{\textwidth}{%
			O.~Reuter, R.~Schubert, M.~Loba, and M.~Maurer, ``Equalized Coverage in Motion Control Performance Prediction for Self-Adaptive Road Vehicles,''
			in \emph{2026 IEEE 29th International Conference on Intelligent Transportation Systems (ITSC)}, Naples, Italy, September 15-18, 2026, {to be published}.
		}%
	}
	\vspace{2cm}
		
\end{@twocolumnfalse}
]
	
\noindent\begin{minipage}{\textwidth}

\hologo{BibTeX}:
\footnotesize
\begin{lstlisting}[frame=single, breakatwhitespace=true]
@inproceedings{reuter_equalized_2026,
    author = {Reuter, Ole and Schubert, Richard and Loba, Marvin and Maurer, Markus},
    booktitle = {2026 IEEE 29th International Conference on Intelligent Transportation Systems (ITSC)},
    title = {Equalized Coverage in Motion Control Performance Prediction for Self-Adaptive Road Vehicles},
    address = {Naples, Italy},
    year = {2026},
    publisher = {IEEE, to be published}
}
\end{lstlisting}
\end{minipage}
\fi

\addtolength{\topmargin}{46pt}

\maketitle

\addtolength{\topmargin}{-32pt} 
\begin{abstract}
Automated driving systems require monitoring mechanisms to ensure operation as intended, especially when system elements degrade and/or fail. Hence, \qemph{capability monitoring} is crucial in order to evaluate the system's remaining performance and implement capability-based behavior. In this paper, we investigate the dynamics of a highly over-actuated automated vehicle under actuator degradations and failures, affecting the vehicle's motion control capabilities. We propose a lightweight prediction model based on \qemph{conformalized quantile regression} that predicts whether an automated vehicle can be controlled with sufficiently low lateral deviation from a planned trajectory under nominal, degraded, and failed actuator conditions. We recognize that statistical guarantees should hold not only across all data (marginal coverage) but also for different regimes within the data (conditional coverage). We therefore employ \qemph{equalized coverage} methods to address this challenge. During runtime behavior generation our predictor can provide a heuristic for determining the admissible action space. Its application and limitations are discussed in this paper.
\end{abstract}
\begin{figure*}[!t]
    \centering
    \vspace{-12pt}
    \includegraphics[width=0.625\textwidth]{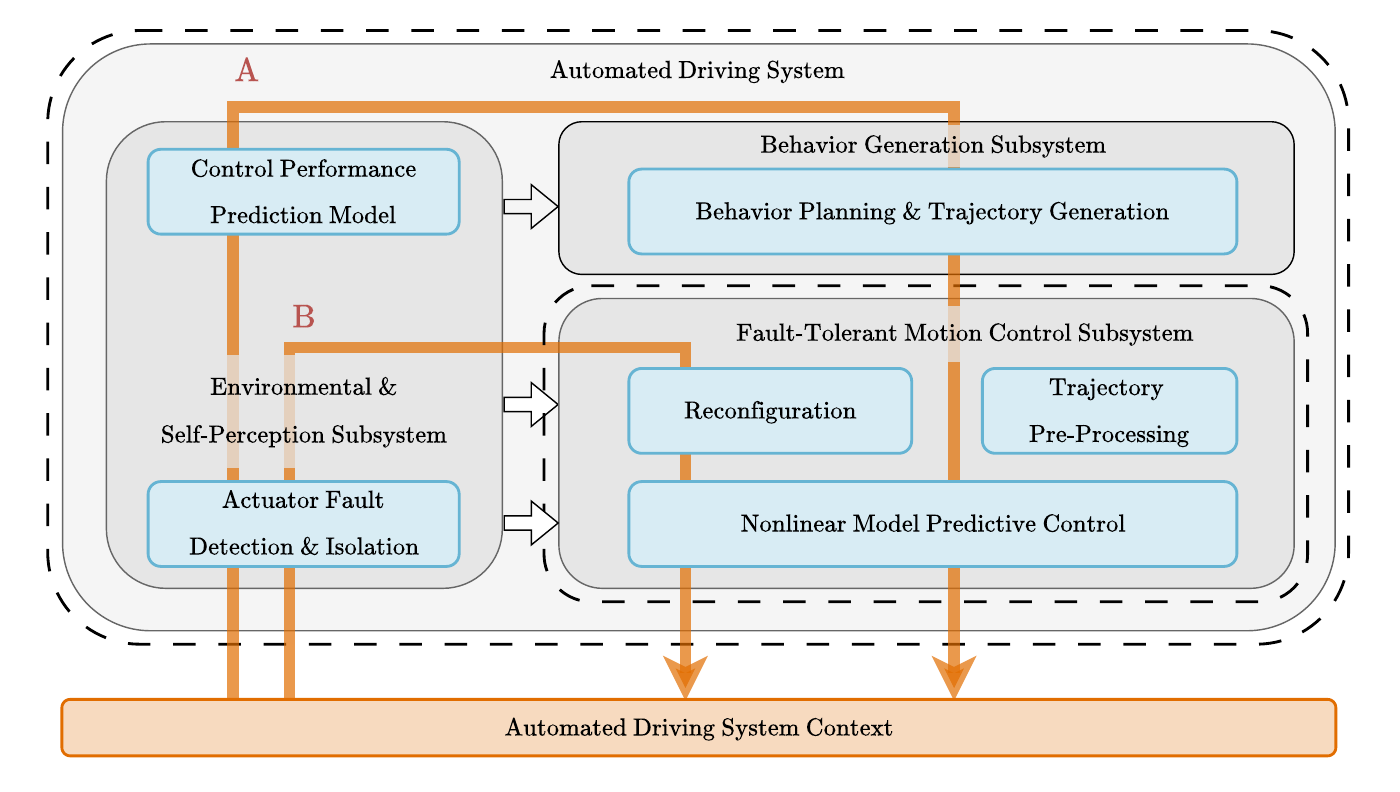}
    \caption{Simplified illustration of a functional architecture taken from~\parencite{schubert_architectural_2026}, based on~\parencite{stolte_faulttolerant_2023, schubert_conformal_2024}, originally adapted from \parencite{ulbrich_towards_2017}. System boundaries of the \ads (outer) and motion control subsystem (inner) are visualized using dashed lines. Control loop like patterns (A, B) are highlighted as orange arrows: (A) On the tactical level, the larger \ads may adapt its behavior in order to generate maneuvers more suitable to the system's current conditions. Both control loops rely on sensor inputs, taken from the context or measured internally, and ultimately affect the system's behavior in its context. (B) Simultaneously, the motion control subsystem may adapt itself internally in order to respond to actor degradations and failures.}
    \label{fig:slx_controlloop} \label{fig:arch}
    \vspace{-0.5cm}
\end{figure*}
\section{Introduction} \label{sec:intro}
Automated driving technology offers the potential to (temporarily) relieve drivers from the tasks of driving and monitoring their vehicle, even in challenging scenarios. To ensure that these systems operate as intended, safety-related constraints on system behavior must be considered during both the design and operational phases. A crucial aspect when designing an automated driving system involves ensuring that the vehicle can maintain operation as intended even in cases of degradation or failure: A \qemph{degradation} represents the state of a system element with reduced performance, yet maintaining given functionality. In contrast, a \qemph{failure} refers to the termination of the system's intended behavior, i.e., the system ceases to provide its functionality~\parencite{stolte_taxonomy_2021}. Beyond robust design, monitoring system performance online is essential~\parencite{nolte_supporting_2020} to detect, isolate, and compensate for degradations and failures before they lead to hazardous behavior of the vehicle.

To enable appropriate adaptation of behavioral decisions at runtime, the system depends on an explicit representation of knowledge about itself, rendering the system \qemph{self-aware}~\parencite{nolte_supporting_2020, schubert_architectural_2026}. Such representation may not be chosen arbitrarily: It must be \qemph{interpretable}~\parencite{schubert_architectural_2026} to promote that system states and actions remain traceable for communication with external (human) reviewers~\parencite{nolte_representing_2018}. At runtime, functional boundaries can be estimated, allowing admissible actions to be determined from such representations and enabling the system to adapt its behavior accordingly---thereby rendering it \qemph{self-adaptive}
~\parencite{schubert_architectural_2026}.

Note that this motivation applies to any kind of automated (driving) system---regardless of its implementation. In particular, both \modarchs (with white-box elements) as well as current \aib and/or even \eendarchs should~\parencite{yuan_endtoend_2024}, in our opinion, provide interpretable representations of system capabilities for traceable behavior generation~\parencite{nolte_representing_2018}, leading to architecture-specific challenges.

This paper extends and revises our previous work~\parencite{schubert_conformal_2024}: We investigate the dynamics of a highly over-actuated automated vehicle under actuator degradations and failures, with a particular emphasis on the explicit representation and systematic analysis of features leading to higher uncertainty therein. We present a lightweight prediction model based on \qemph{conformalized quantile regression}~\parencite{romano_conformalized_2019} that predicts whether an automated vehicle can be controlled with sufficiently low lateral deviation under nominal, degraded, and failed actuator conditions---with statistical guarantees on the estimation of such deviation. In more detail and in extension to~\parencite{schubert_conformal_2024}, we 
\begin{enumerate}
    \item provide method requirements that lead to the chosen method of \CQR,
    \item account for the fact that statistical guarantees should hold not only across all cases (marginal coverage) but for different regimes (conditional coverage). We therefore apply the \qemph{equalized coverage} method by \textcite{romano_malice_2020} and provide guidance on choosing grouping variables for group-equalized coverage,

    \item and finally highlight the limitations of our paper.
\end{enumerate}

The remainder of this paper is structured as follows: \autoref{sec:literature} reviews related work, and \autoref{sec:requirements} outlines the relevant requirements. \autoref{sec:controller} presents the simulation framework, while \autoref{sec:predictor} introduces the conformalized prediction model with equalized coverage. The application is demonstrated in \autoref{sec:application}, and \autoref{sec:discussion} concludes with a critical discussion.

\section{Related Work} \label{sec:literature}
Early ideas of monitoring control loops in hierarchical systems were discussed by \textcite{passino_introduction_1993} and applied to automated driving in~\parencite{maurer_flexible_2000, siedersberger_komponenten_2003}, extending to knowledge-based expert architectures~\parencite{pellkofer_verhaltensentscheidung_2003}. Numerous publications discuss the need to monitor capabilities~\parencite{nolte_supporting_2020} and capability-aware behavior generation~\parencite{nolte_werte_2024} of automated vehicles, contributing to their \qemph{\sawa}~\parencite{schubert_architectural_2026}. %
In particular, \textcite{nolte_supporting_2020} point out that systems require a \qemph{self-representation} to determine their capabilities~\parencite{reschka_fertigkeiten-_2017} and select appropriate actions. This set of actions\footnote{In automated driving, we equate \qemph{action} and \qemph{driving maneuver}, i.e., an abstraction of possible state progressions~\parencite{jatzkowski_zum_2021}.} is referred to as \qemph{admissible action space}~\parencite{nolte_supporting_2020}, with constraints being both \qemph{internal} and \qemph{external}~\parencite{nolte_supporting_2020}.

In~\parencite{nolte_towards_2017}, \citeauthor{nolte_towards_2017} propose applying \qemph{skill graphs}~\parencite{reschka_fertigkeiten-_2017} to identify vehicle capabilities and behavior/capability-level requirements. Similar work is conducted by \textcite{stolte_towards_2020}: Both propose eliciting behavioral requirements during design, addressing quantities related to the vehicle's motion, e.g., maximum lateral deviation from a reference path~\parencite{nolte_towards_2017}. Technical determinants---e.g., state estimation accuracy or actuator performance---influence behavior execution performance. Runtime models (e.g., \qemph{ability graphs}~\parencite{reschka_fertigkeiten-_2017}) are required to monitor system capabilities.

In earlier work, we proposed monitoring the motion control capability of a highly over-actuated automated vehicle~\parencite{schubert_conformal_2024} in the \autotech project~\parencite{van_kempen_autotechagil_2023}.
Similar to our work, \textcite{da_silva_crash-prone_2024} investigate the impact of actuator fault combinations on an over-actuated vehicle's ability to perform evasive lane changes. Their offline analysis employs simulation to acquire performance metrics, including maximum lateral deviation and ``leave road'' and ``collision'' indicators, tracing the impact of degradations and failures---however, only for offline analysis.

As said before, at runtime, vehicles should adapt planned actions to respect current control capabilities. Degradations may cause inconsistencies between planned and actual motion. \textcite{ratner_operating_2023} address ensuring consistency between motion planning and control models for robots, proposing to bias planners away from inaccuracies by increasing costs where discrepancies are observed. For that, stochastic discrepancy models are learned online. A similar approach is presented in~\parencite{noseworthy_active_2021}.
In safety-critical applications, online learning is often infeasible since model discrepancies must be experienced before learning (in other words: before it is too late). \textcite{castaneda_recursively_2023} pair Gaussian Process models with Control Barrier Functions to ensure systems stay within defined bounds under uncertainty, acquiring measurements only when dynamics are expected to slightly deviate from the model.
To evaluate executable actions given actual dynamics, \textcite{mcconachie_learning_2020} propose a neural network classifier predicting whether actions are \inq{reliable} or \inq{unreliable}, where unreliable actions likely violate real-world constraints not considered in simplified planning models. This enables planning in simplified state spaces while filtering unreliable actions before execution.

Ensuring controlled systems stay within defined bounds is also addressed by \qemph{reachability analysis}, used offline as a design tool as well as for online verification~\parencite{althoff_online_2014}. Recent approaches often use \qemph{Hamilton-Jacobi} reachability analysis~\parencite{bansal_hamilton-jacobi_2017, leung_infusing_2020}, but it is computationally expensive and scales exponentially with system states~\parencite{bansal_hamilton-jacobi_2017, chen_high_dimensional_reachability_analysis_2017, leung_infusing_2020, akinwande_strategy_verifying_reach_avoid_2026}. Systems with more than five states are often intractable for online verification, especially for neural network-based models with many latent states. To reduce computational effort, linearized models with few states~\parencite{althoff_online_2014}, sampling-based approximations~\parencite{lew_sampling-based_2021}, and subsystem decomposition~\parencite{chen_decomposition_2018} are used.

Overall, while reachability analysis aims to \qemph{guarantee} state regions are reached/avoided, results depend on assumptions about dynamics models, input spaces, and disturbances. Model-based guarantees should be considered carefully since assumptions may not hold in practice. Sudden degradations and failures (effectively abrupt model changes) require updated dynamics representations, necessitating research on deriving dynamics models under degradations and failures, e.g., as in~\parencite{asadi_damaged_2014} for aviation systems.

Given reservations about \inq{exact} guarantees, approaches providing \qemph{probabilistic} guarantees receive more attention. 
Disturbances are often assumed bounded, but \qemph{stochastic reachability analysis} methods exist~\parencite{vinod_stochastic_2021}. Neural control barrier functions combined with split conformal prediction have been proposed for statistically validated barriers~\parencite{tayal_cp_2025, zhang_conformal_2025}. In general, \qemph{conformal prediction} methods~\parencite{angelopoulos_gentle_2022} are gaining popularity as a lightweight, distribution-free approach for verifying systems, providing statistically valid guarantees for in-distribution problems in motion planning and control~\parencite{lindemann_formal_2025}.

\section{Requirements for Method Selection}\label{sec:requirements}
Building on the given sources above, we aim to select an appropriate method for the problem setting in this paper. Thus, we elicit key method requirements:
\requirement{req:capRuntimeFeasible}{Method Requirement 1}{The method shall exhibit sufficiently low computational cost for online deployment at runtime.}
Methods that are theoretically valid but computationally infeasible for real-time deployment must be excluded. 
For instance, reachability analysis methods typically require exponential computation time depending on the number of system states~\parencite{bansal_hamilton-jacobi_2017, chen_high_dimensional_reachability_analysis_2017, akinwande_strategy_verifying_reach_avoid_2026} (\inq{curse of dimensionality}), limiting their applicability for systems with many states (for the over-actuated vehicle model considered here, a 10-dimensional, nonlinear model is appropriate~\parencite{nolte_sensitivity_2020}).
\requirement{req:capAIBCompatible}{Method Requirement 2}{The method shall remain applicable when AI models are employed for vehicle dynamics modelling and/or control.}
With the increasing adoption of \aib components~\parencite{yuan_endtoend_2024}, even in dynamics modelling and controller design~\parencite{kuuti_survey_2020}, methods for capability monitoring must remain applicable when such techniques are used.
Aforementioned methods for reachability analysis have also been transferred to neural network based systems~\parencite{huang_reachnn_2019, zhang_reachability_2023}. Yet, the curse of dimensionality especially applies to such (potentially) complex models.
\requirement{req:capStatisticalValidity}{Method Requirement 3}{The method shall provide quantitative performance estimates that are interpretable.}
In safety-critical applications, decisions must be made based on incomplete information~\parencite{nolte_werte_2024} but must be traceable in hindsight~\parencite{nolte_representing_2018}. Capability monitoring outputs must hence be interpretable (i.e., yield a geometric, statistical, ... meaning).
This is fulfilled by any method relying (among others) on geometric measures (like most reachability analysis methods) but less by classification states such as \inq{reliable} or \inq{unreliable}~\parencite{mcconachie_learning_2020} that are not straightforward to interpret.
\requirement{req:capStatisticalValidity}{Method Requirement 4}{The method shall quantify prediction uncertainty in a quantitatively and statistically explicit and interpretable manner.}
Given the omnipresence of uncertainty, the method should provide not only nominal performance estimates but also statistically interpretable measures of uncertainty, enabling meaningful assessment of vehicle behavior~\parencite{aravantinos_making_2020}.
The last requirement is satisfied, for instance, by using stochastic models such as Gaussian processes~\parencite{castaneda_recursively_2023} or any form of conformal prediction---with different \inq{backbones} such as Gaussian processes~\parencite{xu_online_2025}, neural networks~\parencite{romano_conformalized_2019} or more advanced analytical methods~\parencite{zhang_conformal_2025}.

When combined with a sparse neural network, conformal prediction meets all requirements by providing distribution-free, statistically valid prediction intervals for arbitrary black-box models. Runtime cost is low, e.g., for a feed-forward neural network as predictor, as in~\parencite{romano_conformalized_2019, romano_malice_2020}. This motivates the approach adopted in~\parencite{schubert_conformal_2024} and refined in this work. To ensure interpretability, we use high-level maneuver parameters as inputs and restrict outputs to geometric measures.
\section{Behavior Generation and Motion Control Framework in Simulation} \label{sec:controller}
In this work, we assume a \modarch for our system, with its behavior generation being split, among others, into a tactical and a stabilization layer. While the former covers the generation of an appropriate maneuver~\parencite{jatzkowski_zum_2021} based on the current situation and system capabilities, the latter takes care of its diligent execution (see also \autoref{fig:slx_controlloop}).

In particular, we examine the performance of the example controller in simulation to build a predictive model thereof to inform runtime behavior generation at the tactical layer. For our case, we choose a fault-tolerant motion controller by~\parencite{stolte_toward_2023} and assess it under diverse dynamic conditions and actuator degradations and failures, extending the approach of~\parencite{schubert_conformal_2024}. 

The controller is deployed in a simulation of \mobile~\parencite{bergmiller_functional_2014}, a highly over-actuated research vehicle with four independently drivable, brakable, and steerable wheels.
The architecture is then implemented in our simulation setup, relying on \qemph{MATLAB} and \qemph{IPG CarMaker}, and described hereafter.

\subsection{Behavior Generation} \label{sec:behavior}
The behavior generation module provides high-level driving maneuvers at runtime by selecting from a discrete set of possible maneuver options $\mathcal{D}_q=\{0, 1, ...\} \ni D_q$ based on (i) the current situation as well as (ii) capability monitoring, i.e., feedback from the prediction of the motion controller's performance. For our case study, we parameterize two lateral maneuvers for changing and following lanes paired with a longitudinal follow speed maneuver via the \qemph{maneuver template}~\parencite{jatzkowski_zum_2021} $\vec{m_q} = (D_q, r_q, v_q, \amax)$ with direction (if any) $r_q \in \{-1, 0, 1\}$. $v_q$ denotes the targeted terminal velocity and $\amax$ the maximum permissible lateral acceleration. We also can encode a \inq{\mrm} to stop the vehicle in the current lane, $\vec{m_q} = (0, 0, 0, \amax)$.

\subsection{Model-Based Trajectory Generation} \label{sec:planner}
Given the maneuver template $\vec{m_q}$, we employ a model predictive controller to plan a corresponding reference trajectory at runtime. In addition, road-geometric aspects are considered, e.g., lane width $w_q(s)$ and lane curvature $k_q(s)$ as a function of (Frenet) arc length $s$. We use a linear parameter-varying single-track model based on~\parencite{nolte_werte_2024} that describes the dynamics of states $\vec{x}=(\beta, \dot\psi, e_\psi, e_d, \delta_f, \delta_r)$ as a function of control inputs $\vec{u}=(\dot\delta_f, \dot\delta_r)$ and disturbance $\dot\psi_{\mathrm{ref}}$. Here, $\beta$ denotes the sideslip angle, $\psi$ is the yaw angle, $e_\psi$ and $e_d$ describe the angular and lateral offset from the desired path and $\delta_{\{f,r\}}$ denote the steering angle at the model's front and rear wheel. The additive disturbance $\dot\psi_{\mathrm{ref}}$ describing the reference path's yaw rate is calculated using the lane's curvature and the velocity.
For maneuver settings with varying speed, a simple PID controller can be additionally employed (which becomes particularly crucial for the \inq{\mrm} in \autoref{sec:application}).
See also~\parencite{nolte_model_2017, nolte_werte_2024} for details.

\subsection{Fault-Tolerant Trajectory Tracking Controller} \label{sec:FTC}
We apply the generated reference trajectory to the trajectory tracking controller by~\parencite{stolte_toward_2023} that uses a nonlinear model predictive control approach. Relying on a nonlinear double-track model paired with an adapted version of the Fiala tire model~\parencite{hindiyeh_controller_2014}, the MPC employs control inputs $\vec{u}_{\mathrm{MPC}} = (\tau_{\wheel}, \dot{\delta}_{\wheel})$ to control the states $\vec{x}_{\mathrm{MPC}} = (s, d, \psi, v_x^\mathrm{V}, v_y^\mathrm{V}, \dot{\psi}^\mathrm{V}, \delta_{\wheel})$ with wheels $w \in \{\mathrm{fl, fr, rl, rr}\}$. Here, $\tau_{\wheel}$ is the wheel torque and $\delta_{\wheel}, \dot{\delta}_{\wheel}$ are the steering angle and rate. $s$ is the Frenet arc length of the reference trajectory, $d$ denotes the lateral deviation, $\psi$ is the yaw angle, and $v_x^\mathrm{V}, v_y^\mathrm{V}$ are the longitudinal and lateral velocities in the vehicle frame.

The trajectory tracking controller is actively reconfigured based on current actuator degradations or failures~\parencite{stolte_toward_2023}, representing a form of internal adaption as highlighted in \autoref{fig:arch}. To mathematically represent deviations from the fault-free case, we adjust the permissible bounds on each actuator variable $z\in \{\delta, \dot{\delta}, \tau\}$ to obtain a percentage of remaining performance: Let $z_{\wheel}^{\min}$ and $z_{\wheel}^{\max}$ denote the nominal actuator limits and let $[z_{\degradation, \wheel}^{\min},z_{\degradation, \wheel}^{\max}]$ be the degraded value range. By using $|z|_{\degradation,\wheel} := \max\{|z_{\degradation, \wheel}^{\min}|,|z_{\degradation, \wheel}^{\max}|\}$, we then define $\zdegwheel:=\frac{|z|_{\degradation,\wheel}}{z_{\wheel}^{\max}} \in [0, 1]$. See~\parencite{schubert_conformal_2024} for details.

\section{Conformalized Prediction Model} \label{sec:predictor}
At runtime, we wish to identify maneuver parameters that yield a sufficiently low control deviation---reflecting the reconfigured controller's capability to compensate for actuator degradations. Therefore, we explore a finite set of maneuver parameters, exploiting the availability of environmental information as well as fault detection and isolation data.

\subsection{Dataset Creation} \label{sec:dataset}
To learn a prediction model, we construct the required data. Our dataset is created using real-world road geometries measured along the inner-city ring road of \textit{Braunschweig, Germany}. We obtain $N_{\mathrm{S}}=222$ road segments and, for each segment, generate $N_{\mathrm{M}}=15$ maneuver templates (lane change left/right or lane follow) with varying speeds and maximum lateral accelerations (five candidate acceleration levels, speed randomly selected for each run). To capture the influence of actuator degradations and failures, we create $N_{\mathrm{D}}=10$ degradation settings (nominal case and nine randomly degraded cases, affecting values of $\zdegwheel$). This results in a total of $N_\mathrm{S} \cdot N_\mathrm{M} \cdot N_\mathrm{D} = 33300$ simulation runs, for each of which we record the maximum lateral deviation $\epsNmax$.
To summarize utterly infeasible cases and/or such that yield drastic deviations, we abort the simulation and clip the measured maximum lateral deviation. The cutoff value is set to $\epsCutoff := 0.675\meters$, corresponding to the available lateral clearance on each side of the vehicle, assuming an average lane width of $\bar{w} = 3.31\meters$ and a vehicle width of $1.96\meters$. 
The complete set of simulation input parameters together with the ranges observed in the generated dataset are summarized in~\autoref{tab:mlp_inputs_outputs}.
\begin{table}%
    \centering
    \vspace{0.5em}
    \caption{Variables gathered as simulation output and used to train the predictor, with input $\inputnn$ and output $\outputnn = \epsNmax$.\protect\footnotemark}
    \begin{adjustbox}{center, keepaspectratio, width=0.95\columnwidth}
        \renewcommand{\arraystretch}{1.25}
        \begin{tabular}{cccc}
            \toprule
            \textbf{Variable} & \textbf{Description} & \textbf{Unit} & \textbf{Value Ranges in Data} \\
            \midrule
            $r_{\odd}$ & direction & $1$ & $\{-1, 0, 1\}$ \\
            $\wmax$ & max. lane width & $\tm$ & $[2.62,\, 5,67]$ \\
            $\kmin$ & min. curvature & $\toom$ & $[-3.5\cdot 10^{-2},\, 0]$ \\
            $\kmax$ & max. curvature & $\toom$ & $[0,\, 4.0\cdot 10^{-2}]$ \\
            $v_{\odd}$ & velocity & $\tkmh$ & $[30,\, 50]$ \\
            $\amax$ & max. acceleration & $\tmss$ & $[2.5,\, 4.5]$ \\
            $\deltadegwheel$ & steering angle factor & $1$ & $[0,\, 1]$ \\
            $\deltadotdegwheel$ & steering rate factor & $1$ & $[0,\, 1]$ \\
            $\taudegwheel$ & wheel torque factor& $1$ & $[0,\, 1]$ \\
            \addlinespace
            \midrule
            $\outputnn = \epsNmax$ & max. lateral deviation & $\tm$ & $[0.0,\, 1.02]$ \\
            \bottomrule
        \end{tabular}
        \end{adjustbox}
    \label{tab:mlp_inputs_outputs}
\end{table}

\subsection{Conformalized Quantile Regression}
Based on the obtained data, we aim to learn the mapping  
\begin{equation}
	y = f\bigl(\wmax, \kmax, \dots ,
	\dot{\delta}_{\mathrm{D,\wheel}},
	\tau_{\mathrm{D,\wheel}}\bigr)
	\;=\; f(\inputnn) ,
	\label{eq:target_function}
\end{equation}
where $\inputnn$ denotes the input vector (see \autoref{tab:mlp_inputs_outputs}) and the output $\outputnn$ is the maximal lateral deviation $y = \epsNmax$. In this work, we approximate $f$ with a \emph{neural network} that learns the function
\begin{equation}
    \hat{y}= \hat{f}(\inputnn) \approx f(\inputnn).
\end{equation}
We employ \emph{quantile regression}~\parencite{koenker_regression_1978}, which, rather than delivering a single point estimate $\hat{y}$, provides a prediction interval
$\interval(\inputnntest) := [\hat{y}_{\sulow},\, \hat{y}_{\suup}]$
for any test sample $\inputnntest\in\inputnntestset$. For constructing the intervals, we adopt \emph{\CQR} (CQR)~\parencite{romano_conformalized_2019, angelopoulos_gentle_2022}, which guarantees that, for a pre-specified \inq{error rate} $\alpha$, the interval coverage satisfies  
\begin{equation} \label{eq:conformal_prediction}
	    \prob\left \{\outputnntest \in \interval(\inputnntest) = [\hat{y}_{\sulow},\, \hat{y}_{\suup}] \right \} \geq 1-\alpha =: \coverage.
\end{equation}

\footnotetext{The minimum lane width is recorded as well to allow for a conservative estimation of available free space. In order to generate conservative predictions, we solely use the maximum lane width as an input for the predictor.}

\subsection{Equalized Coverage}
\label{sec:eqcoverage}
Equation~\eqref{eq:conformal_prediction} ensures \emph{marginal coverage}: on average, a prediction interval contains the true target with probability at least $\coverage$ across all test samples. However, this guarantee may not hold for specific subsets, such as certain degradation levels, maneuver types, or road-segment characteristics. We argue that prediction intervals must provide reliable uncertainty estimates across all operating conditions, avoiding systematic over- or under-coverage.

To enforce this property, we extend the approach from~\parencite{schubert_conformal_2024} to achieve \emph{equalized coverage}~\parencite{romano_malice_2020}: Let $G$ denote a set of distinct groups that partition the data. For each group $g$, we require that the prediction interval $\interval(\inputnntest,G)$ satisfies
\begin{equation} \label{eq:equalized_coverage}
    \prob\left \{\outputnntest \in \interval(\inputnntest, G) \mid G = g \right \} \geq \coverage.
\end{equation}

In particular, we follow the equalized coverage procedure of group-conditional conformalization~\parencite{romano_malice_2020} that is rather data-efficient: The quantile regression model is trained on the full training set, while the calibration step is carried out separately for each subgroup using only the calibration samples belonging to that group.

\subsection{Preparation of Groups for Equalized Coverage}
As defined, equalized coverage requires the data to be split into $2,\dots,N_G$ groups. In the simplest case, we can define a single threshold variable that splits it in half. For selecting a meaningful grouping variable, however, we find no methodological guidance in~\parencite{romano_malice_2020}.

\begin{table}%
    \centering
    \caption{Feature scores based on Mutual Information, MRMR selection rank, and Breusch-Pagan F-statistic. Bold formatting denotes the top-ranked features within each feature evaluation method.}
    \begin{adjustbox}{center, keepaspectratio, width=0.95\columnwidth}
        \renewcommand{\arraystretch}{1.35}
        \begin{tabular}{cccc}
            \toprule
            \textbf{Variable} & \textbf{Mutual Information} & \textbf{MRMR} & \textbf{Breusch-Pagan} \\
            \midrule
            $w_{\mathrm{max}}$ & \textbf{0.669} & Rank 14 & $\approx$ 0 \\
            $v_{\odd}$ & \textbf{0.212} & Rank 4 & 2.5 \\
            $a_{\mathrm{max}}$ & 0.098 & \textbf{Rank 3} & 51.8 \\
            $\tilde{\delta}_{\mathrm{D,fl}}$ & 0.126 & Rank 5 & \textbf{982} \\
            $\tilde{\delta}_{\mathrm{D,fr}}$ & 0.127 & Rank 6 & \textbf{920} \\
            $\tilde{\delta}_{\mathrm{D,rl}}$ & 0.120 & Rank 11 & 479 \\
            $\tilde{\delta}_{\mathrm{D,rr}}$ & 0.119 & Rank 10 & 503 \\
            $\dot{\tilde{\delta}}_{\mathrm{D,fl}}$ & 0.121 & Rank 8 & 828 \\
            $\dot{\tilde{\delta}}_{\mathrm{D,fr}}$ & 0.126 & Rank 9 & 754 \\
            $\dot{\tilde{\delta}}_{\mathrm{D,rl}}$ & 0.122 & Rank 12 & 544 \\
            $\dot{\tilde{\delta}}_{\mathrm{D,rr}}$ & 0.128 & Rank 13 & 482 \\
            $|k|^{\max}$ & \textbf{0.727} & \textbf{Rank 1} & 370 \\
            $D_{\geq 2}^{(0.1)}$ & 0.050 & \textbf{Rank 2} & \textbf{2685} \\
            $D_{\geq 2}^{(0.2)}$ & 0.046 & Rank 7 & \textbf{1647} \\
            \bottomrule
        \end{tabular}
        \end{adjustbox}
    \label{tab:feature_importance}
\end{table}

As a first step towards the solution, feature selection techniques such as filter methods could estimate the impact of a feature (or determinant) on the target variable~\parencite{guyon2006introduction}. However, equalized coverage is used to tackle the \qemph{heteroscedasticity} of the data. It is hence more about identifying variables that lead to higher \qemph{uncertainty} than absolute impact (rendering typical methods like linear correlation rather useless).
While this might sound counter-intuitive at first, we state that dynamic and degradation variables with assumed high impact on motion control performance~\parencite{schubert_odd-centric_2023, da_silva_crash-prone_2024} are not necessarily good \qemph{grouping} variables.
Therefore, using measures like \qemph{Mutual Information} or methods like \qemph{Maximum Relevance and Minimum Redundancy} (MRMR) building on it~\parencite{zhao2019mrmr_marketing_platform, smazzanti2024mrmr_selection} is more reasonable, e.g., than estimating linear correlation. Also, statistical tests for heteroscedasticity (like Breusch-Pagan~\parencite{breusch1979heteroscedasticity_test}) can be applied. Results are shown in \autoref{tab:feature_importance}.\footnote{The Brown-Forsythe test~\parencite{brown1974robust} for variance equality was also conducted in the background and shows similar patterns to Breusch-Pagan.} Note that the importance scores for wheel torque degradation variables $\taudegwheel$ (not considered here) are negligible across all methods, which meets our intuition as we consider lateral deviations.

Especially for steering-angle-related degradation variables, the influence seems to be highly nonlinear, as continuous degradation variables yield not necessarily high feature importance scores but additionally designed conditional dummy variables $D_{\geq N_{\mathrm{W}}}^{(\ell_{\mathrm{D}})}$ do, following a simple condition like \inq{set to $1$ if at least $N_{\mathrm{W}}$ degradation parameters yield values $\leq \ell_{\mathrm{D}}$}\footnote{We define
\begin{equation*}
    D_{\geq N_{\mathrm{W}}}^{(\ell_{\mathrm{D}})} := \mathbb{I}\!\left(\sum_{w}\sum_{z} \mathbb{I}(\tilde{z}_{\mathrm{D},w} \leq \ell_{\mathrm{D}}) \geq N_{\mathrm{W}}\right).
\end{equation*}}.
Among selected methods, curvature measure 
\begin{equation}
|k|^{\max} = \max(|k_{\odd}^{\min}|,|k_{\odd}^{\max}|)
\end{equation}
ranks highest across mutual information and MRMR methods, indicating strong predictive power. $D_{\geq N_{\mathrm{W}}}^{(\ell_{\mathrm{D}})}$ dummies rank highest (for low $\ell_{\mathrm{D}}$) in heteroscedasticity-specific tests (Breusch-Pagan). Also, (continuous) front-wheel degradation variables show higher importance relative to rear-wheel degradation variables.
Surely, with each of these \inq{important} variables, several assumptions are introduced. Especially geometric properties are condensed into scalar indicators. Accordingly, the considered variables are \inq{loaded} with high uncertainty---which the analysis proofs.

Based on the obtained results, we (hyper-)train the neural network iteratively and find that we obtain the best results using a binary grouping variable defined as (with $k_{\mathrm{thresh}} = 0.003/\mathrm{m}$)
\begin{equation}
    D_{\mathrm{group}} := |k|^{\max} > k_{\mathrm{thresh}}.
\end{equation}

\subsection{Model Training and Evaluation}\label{sec:training_evaluation}
From the overall data, a testing set with $|\inputnntestset| = 4000$ is held out. From the remaining data, we define a calibration set $\inputnncalset$ with $|\inputnncalset| = 4000$ for the conformal prediction procedure. The remaining data $\inputnntrainset$ is used for training.

\begin{table}
    \centering
    \caption{Network Architecture and Training Parameters}
    \renewcommand{\arraystretch}{1.2}
    \begin{adjustbox}{width=0.94\linewidth}
        \begin{minipage}[t]{0.55\linewidth}
            \centering
            \begin{tabular}{ccc}
                \toprule
                \textbf{Layer} & \textbf{Neurons} & \textbf{Activation} \\
                \midrule
                Linear 1 & 19 & ReLU \\
                \rule{0pt}{4.08ex} %
                Linear 2 & 380 & ReLU \\
                \rule{0pt}{4.08ex} %
                Linear 3 & 380 & ReLU \\
                \rule{0pt}{4.08ex} %
                Linear 4 & 2 & ReLU \\
                \bottomrule
            \end{tabular}
        \end{minipage}%
        \begin{minipage}[t]{0.025\linewidth}
            \centering
            \vspace{0pt}
        \end{minipage}%
        \begin{minipage}[t]{0.55\linewidth}
            \centering
            \begin{tabular}{cc}
                \toprule
                \textbf{Parameter} & \textbf{Value} \\
                \midrule
                Batch Size & 128 \\
                Optimizer & Adam \\
                Learning Rate & 0.0005 \\
                Num. of Epochs & 1200 \\
                Early Stopping & True \\
                Normalization & Batch Norm \\
                \bottomrule
            \end{tabular}
        \end{minipage}
    \end{adjustbox}
    \label{tab:mlp_arch}
\end{table}
The predictor is trained using the framework introduced in~\parencite{romano_conformalized_2019, romano_malice_2020}\footnote{{\smallerfootnote\url{https://github.com/yromano/cqr}, Accessed: Jan. 9, 2026.}}. To optimize the neural network training, we perform a grid search over the most influential hyperparameters of the conformal prediction model. The search space was constructed on the basis of preliminary experiments and experience from prior work. A selection of the most influential parameters and their chosen values is shown in~\autoref{tab:mlp_arch}.
To rank the resulting predictors, we first check for conformance with the targeted coverage on $\inputnntestset$ with tolerance $\Delta C_{\alpha}=\pm 1\%$. From the remaining predictors, we evaluate the $90^{\mathrm{th}}$ percentile of resulting interval lengths and chose the predictor yielding the shortest intervals.

\begin{figure}[htb]
	\centering
	\includegraphics[width=0.95\columnwidth]{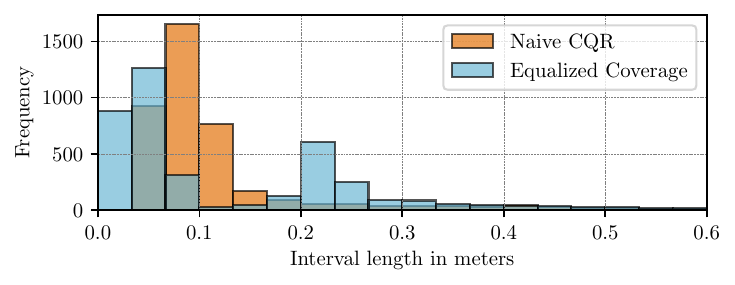}
	\caption{Interval length distributions for predictions on $\inputnntestset$, with and without the equalized coverage method.}
	\label{fig:compare_intervalLenghts}
\end{figure}

To further investigate the impact of using equalized coverage methods, we train a similar predictor using the marginal ("naive") CQR routine.
As shown in \autoref{fig:compare_intervalLenghts}, calibration for equalized coverage does not necessarily produce shorter prediction intervals, but it achieves a more balanced distribution of coverage across data subsets. 
Intuitively, intervals can be shorter for groups with segments exhibiting lower curvatures, whereas higher curvature—reflecting greater geometric variability—induces increased uncertainty and therefore requires wider prediction intervals (see \autoref{sec:discussion}).

Comparing the empirical coverage levels on $\inputnntestset$ with the targeted level of $C_\alpha = 90\%$, we obtain $C_{\alpha,\mathrm{test}}^{\mathrm{eqCov}} = 90.6\%$ for the predictor calibrated for equalized coverage, and $C_{\alpha,\mathrm{test}}^{\mathrm{naive}} = 90.8\%$ for the naive CQR procedure. Hence, both methods achieve marginal coverage close to the nominal level on the test set.
When dividing the evaluation according to the previously defined groups based on lane curvature, differences between the approaches become apparent. The naive CQR method yields $C_{\alpha,\mathrm{test,1}}^{\mathrm{naive}} = 94.3\%$ for lanes with lower curvature and $C_{\alpha,\mathrm{test,2}}^{\mathrm{naive}} = 84.8\%$ for lanes with higher curvature.
For the predictor calibrated for equalized coverage, the corresponding group-wise coverage levels are $C_{\alpha,\mathrm{test,1}}^{\mathrm{eqCov}} = 90.5\%$ and $C_{\alpha,\mathrm{test,2}}^{\mathrm{eqCov}} = 90.8\%$. These values indicate a closer alignment of coverage across the two curvature-based groups while maintaining overall coverage near the target level.

\section{Example Application} \label{sec:application}
\begin{table}
	\centering
	\vspace{0.35em}
	\caption{Degradations examined in the application scenario. Coefficients describing degraded ranges of wheel torques are omitted for the sake of brevity.}
	\begin{adjustbox}{center, keepaspectratio, width=\columnwidth}
		\renewcommand{\arraystretch}{1.2}
		\begin{tabular}{lcccccccc}
			\toprule
			$\zdegwheel [1]$ & $\delta_\mathrm{fl}$ & $\delta_\mathrm{fr}$ & $\delta_\mathrm{rl}$ & $\delta_\mathrm{rr}$ & $\dot\delta_\mathrm{fl}$ & $\dot\delta_\mathrm{fr}$ & $\dot\delta_\mathrm{rl}$ & $\dot\delta_\mathrm{rr}$ \\ 
			\midrule
			D0 & $1$ & $1$ & $1$ & $1$ & $1$ & $1$ & $1$ & $1$\\
			D1 & $0.54$ & $0.44$ & $0.70$ & $0.36$ & $0.87$ & $0.75$ & $0.50$ & $0.55$\\
			D2 & $0.32$ & $0.15$ & $0.59$ & $0.08$ & $0.12$ & $0.02$ & $0.47$ & $0.25$\\
			\bottomrule
		\end{tabular}
	\end{adjustbox}
	\label{tab:app_deg}
\end{table}
The general application of the predictor is also illustrated in~\parencite{schubert_conformal_2024}, so only a brief summary is provided here.

\begin{figure*}[!t]
	\centering
	\begin{subfigure}[t]{0.45\textwidth}
		\centering
		\vspace{-12pt}
		\includegraphics[width=\linewidth]{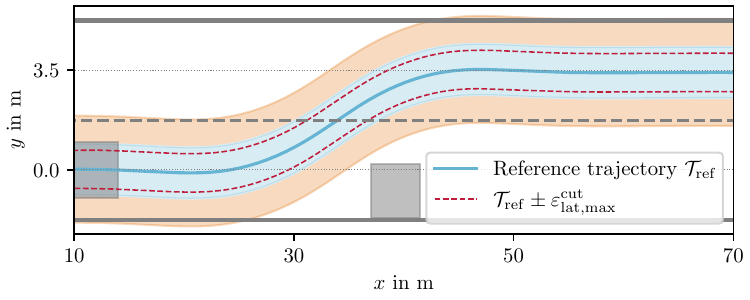}
		\caption{Reference trajectory with the associated lateral deviation envelope (blue shaded area). If the blue envelope exceeds the admissible lateral deviation threshold (red dashed), the maneuver is classified as infeasible, since resulting possible vehicle footprints (orange shaded area) extend beyond the lane boundaries in such cases.}
		\label{fig:traj_buffer}
	\end{subfigure}
	\hfill
	\begin{subfigure}[t]{0.45\textwidth}
	\centering
	\vspace{-12pt}
	\includegraphics[width=\linewidth]{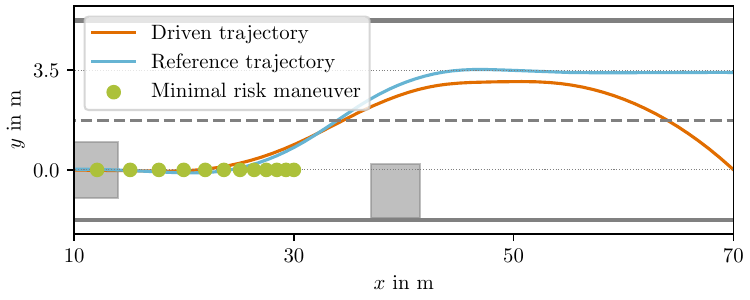}
	\caption{Visualization of an exemplary \mrm method for scenarios, in which all candidate maneuvers are rejected due to high predicted lateral deviations.}
	\label{fig:mrm}
	\end{subfigure}
	\caption{Visualization of the application scenario for degradation D2 and $\amax=4.5\mss$.}
	\label{fig:app}
	\centering
    \vspace{-0.5cm}
\end{figure*}

We consider the following simple scenario: The ego vehicle enters a two-lane road segment containing an obstacle in its lane (e.g., a parked vehicle). Consequently, the behavior generation module faces two alternatives: (i) either executing a lane change into the adjacent lane while selecting an appropriate maximum lateral acceleration or (ii) remaining in the current lane and stopping in front of the obstacle.
We analyze the exemplary straight lane segment shown in \autoref{fig:app} (chosen for easy display), which features a constant lane width of $\wmin = \wmax \approx3.47\meters$. %

Assuming a typical German inner-city driving speed of $v_{\odd}=50\kmh$ and a discrete set of candidate maximum lateral accelerations $\amax=\{2.5, 3,\dots4.5\}\mss$ considered by the predictor, we evaluate feasible maneuver choices assuming different degradation settings as listed in \autoref{tab:app_deg}. D0 represents the fault-free case, whereas D1 and D2 define one slightly and one heavily degraded setup.

Applying the conformalized prediction algorithm yields the predicted upper bounds on the lateral deviation shown in \autoref{tab:app_pred}, representing the deviations the vehicle is expected to produce when attempting to track a trajectory planned for the given maneuver. Given the current lane's minimum width $w_q^{\mathrm{min}}$ and the vehicle width $w_{\mathrm{veh}} = 1.96\meters$, a conservative lateral clearance of $0.5\cdot(w_q^{\mathrm{min}} - w_{\mathrm{veh}}) \approx 0.75\meters$ per side is assumed over the entire segment. A maneuver is rejected if the predicted lateral deviation (i) exceeds this available clearance or (ii) reaches the cutoff value $\epsCutoff$.

As illustrated in \autoref{fig:traj_buffer}, a planned trajectory is expanded by the predicted lateral deviation (blue shaded area). If this expanded region exceeds the cutoff threshold, the maneuver is classified as infeasible; if executed, the resulting uncertainty bounds (orange shaded area) would extend beyond the target lane boundaries. If all candidate maneuvers are rejected, a \inq{\mrm} (e.g., controlled deceleration to a stop within the current lane) is initiated, as shown in \autoref{fig:mrm}.
\begin{table}
	\centering
	\vspace{0.35em}
	\caption{Comparison of measured maximum lateral control errors $\varepsilon_{\mathrm{lat,max}}$ and their predicted upper bounds $\hat{\varepsilon}_{\mathrm{lat,max}}$ for varying values of $\amax$\protect\footnotemark. Maneuvers selected by behavior generation are highlighted in bold.}
	\renewcommand{\arraystretch}{1.2}
		\centering
		\begin{tabular}{lccccc}
			\toprule
			$a_\mathrm{max} [\tmss]$ & $2.5$ & $3$ & $3.5$ & $4$ & $4.5$ \\
			\midrule
			$\varepsilon_\mathrm{lat,max,D0} [\meters]$ & $0.19$ & $0.21$ & $0.23$ & $0.25$ & $\mathbf{0.27}$\\
			$\hat\varepsilon_\mathrm{lat,max,D0} [\meters]$ & $0.23$ & $0.25$ & $0.28$ & $0.30$ & $\mathbf{0.32}$\\
			\midrule
			$\varepsilon_\mathrm{lat,max,D1} [\meters]$ & $0.19$ & $0.2$ & $0.23$ & $0.26$ & $\mathbf{0.28}$\\
			$\hat\varepsilon_\mathrm{lat,max,D1} [\meters]$ & $0.22$ & $ 0.24$ & $0.26$ & $0.29$ & $\mathbf{0.31}$\\
			\midrule
			$\varepsilon_\mathrm{lat,max,D2} [\meters]$ & $0.26$ & $0.29$ & $0.32$ & $\mathbf{0.35}$ & $0.68$\\
			$\hat\varepsilon_\mathrm{lat,max,D2} [\meters]$ & $0.45$ & $0.487$ & $0.49$ & $\mathbf{0.56}$ & $0.87$\\
			\bottomrule
		\end{tabular}
	\label{tab:app_pred}
\end{table}
\footnotetext{Note, that the measured lateral deviations are clipped to a maximum of $0.675\meters$ in the captured datasets, see \autoref{sec:dataset} for details.}
\section{Critical Discussion and Future Work} \label{sec:discussion}
While this example highlights the overall applicability, we take a more critical look on the approach used. In addition to further limitations highlighted in~\parencite{schubert_conformal_2024}, assumptions play a critical role: This work focuses on interpretable representations of the system's capabilities that are easily understood. To obtain such representations, several additional assumptions were introduced beyond those inherent to the simulation model. These include specific formulations of maneuvers as well as selected dynamical and road geometric parameters. Together, these choices abstract from and simplify certain real properties.

Luckily, conformal prediction can compensate the effect of most of these assumptions: Conformalized prediction intervals stay valid regardless of the predictors design~\parencite{angelopoulos_gentle_2022}, i.e., including the assumptions that went in. If we pick any maneuver to be executed (on any road geometry, given any dynamic state, etc.: As long it is an \inq{in-sample}) and we derive input features for the predictor using the methodology described in this paper consistently, the uncertainty introduced in the assumptions is \qemph{explicitly} reflected in the interval lengths.
As a summary, we note that conformal prediction can capture and intuitively represent uncertainty introduced through
\begin{itemize}
    \item model assumptions in the model-based behavior generation stack (versus real or assumed-to-be-real simulation world)
    \item simplifying assumptions in maneuver definitions and the system's dynamical parameters,
    \item road-geometric parameter assumptions, and
    \item data-based model learning.
\end{itemize}

However, it is crucial to emphasize that several limitations remain beyond the methodological scope of this study. Most notably, the primary limitation of the present work lies in the comparatively narrow scope of the proposed method.
Predicting an expected maximum deviation is helpful for maneuver decisions as described. Still, the choice of an admissible level of such deviation is subject to further assumptions. In this paper's example scenario, we assumed a stationary object that the ego-vehicle shall go around. Dynamic objects, though---like moving pedestrians, maybe even behind an occlusion---would change the setting drastically: What walking speed can we assume for the pedestrian? What \emph{safe} lateral distance should be chosen? (See also~\parencite{nolte_representing_2018, graubohm2023assumptions}.)
Accordingly, we note that our method \qemph{cannot} support making statements about, for instance,
\begin{itemize}
    \item out-of-distribution settings and/or
    \item any other system task/capability apart from estimating motion control deviation, especially not accuracy of
    \begin{itemize}
        \item the motion planning underlying motion control,
        \item dynamic object perception, and/or
        \item behavior anticipation.
    \end{itemize}
\end{itemize}

Obviously, methods for complex motion prediction exist, even incorporating the method of conformal prediction~\parencite{doula2023conformal, lindemann2023safe, liang2024safe}. Yet, both expert-based thresholds and more sophisticated, dynamic and often data-based prediction methods rely on expert- and/or data-based assumptions. Highlighting those is critical for transparent communication towards people exposed to the ego-vehicle's actions~\parencite{nolte_werte_2024}. These aspects are subject to the uncertainty introduced through the environment (the \qemph{open context}~\parencite{nolte_werte_2024})---not a sign of a poorly chosen methodology.

A key limitation of this study is the reliance on synthetic simulation data. While simulation enables systematic, scalable, and cost-effective exploration of rare, impractical, and/or even dangerous scenarios that would be infeasible to (re-)produce experimentally, it inevitably introduces modeling assumptions through the simulation software. Consequently, the validity and reliability of the predictor in real-world operation cannot be guaranteed without empirical validation. Future work must therefore focus on the latter.

Finally, and coming back to our introduction, we acknowledge current advancements in the field of \aibss and specifically \eendarchs. Therefore, applying our method to behavior generation and/or control models learned from data would be highly interesting. As we have stated that our method can operate regardless of the \inq{black box} controller's inner workings, we are looking forward to seeing future work on the application of our capability monitoring approach to \aibss and its benefits.

\ifisanonymized
	\rule{\dimexpr\linewidth}{\AnonymizedAcknowledgementHeight}
\else
	\section{Acknowledgement}
We thank Jens Rieken for providing the road segment data.
\fi

\printbibliography

\end{document}